\documentclass[a4paper]{jpconf}
\usepackage{graphicx}
\usepackage{epsfig}
\usepackage{epstopdf}
\begin{document}
\title{Two particle correlations: a probe of the LHC QCD medium}

\author{{\bf Yaxian Mao}$^{1,2}$, for the ALICE
Collaboration}

\address {{$^{1}$ Laboratoire de Physique Subatomique et de Cosmologie, Grenoble 38026, France}

{$^{2}$ Key Laboratory of Quark \& Lepton Physics~(Huazhong Normal
University),  Ministry of Education, Wuhan 430079, China}}

\ead{maoyx@iopp.ccnu.edu.cn}

\begin{abstract}
The properties of $\gamma$--jet pairs  emitted in heavy-ion
collisions provide an accurate mean to perform a tomographic
measurement of the medium created in the collision through the study
of the medium modified jet properties. The idea is to measure
the distribution of hadrons emitted on the opposite side of the 
direct photon.
The feasibility of such measurements is studied by applying the approach on the simulation data, we have demonstrated that this method allows us to measure, with a good
approximation, both the jet fragmentation and the back-to-back
azimuthal alignment of the direct photon and the jet. Comparing
these two observables measured in pp collisions with the ones
measured in AA collisions reveals the modifications induced by the
medium on the jet structure and consequently allows us to infer
the medium properties. 
In this contribution, we discuss a first attempt of such measurements applied to real proton-proton data from the ALICE experiment.
\end{abstract}

\section{Introduction}
Quantum ChromoDynamics (QCD)~\cite{QCD} is a theory of the strong
interaction, the fundamental force describing the interactions of
quarks and gluons making up hadrons. The QCD calculations 
performed on a lattice,
indicate that a phase transition from normal hadronic matter to partonic matter, the Quark-Gluon Plasma (QGP), will occur beyond a critical temperature of $T_{\rm c} \sim~170$~MeV~\cite{phase}.  By colliding heavy ions at ultra relativistic energies,
this new state of matter can be created and
its properties, such as the equation of state, the degrees of freedom and the transport properties can be measured.

The phase diagram has been explored in various regions with heavy-ion collisions at continuously increasing
kinetic energies. Experiments at CERN's Super Proton Synchrotron
(SPS)~\cite{SPS} concluded
on the indirect evidence of a "new state of matter". Current
experiments at Brookhaven National Laboratory's Relativistic Heavy
Ion Collider (RHIC)~\cite{RHIC} have found that matter 
does not behave as an ideal gas of free quarks and gluons
predicted by theory, but, rather, as an almost perfect
fluid. The new experiment ALICE~\cite{ALICE} at
CERN's Large Hadron Collider (LHC), will push further the
study of the QCD medium.
Thanks to the huge step in collision energy ($\sqrt{s_{NN}} = 5.5~TeV$ in Pb-Pb collisions),
LHC will open new avenues for the exploration of matter under extreme conditions of temperature and density.
Since the hot QCD medium will be formed at higher temperatures than at RHIC, the deconfined phase will last longer and more readily modify our experimental probes, allowing for a more accurate study of this new state matter.

Hard scattered partons produced in 
initial stage of the collisions, 
have been identified as a valuable probe of the medium. Indeed,
medium properties can be inferred from the modifications experienced
by the partonic shower inside the medium. Partons are only observed
indirectly, as a collimated jet of hadrons coming from the fragmentation of the partonic shower~\cite{FF}.
Comparing the measurements of the jet fragmentation in proton-proton and heavy-ion collisions will reveal the modifications produced by the medium on the hard scattered partons.
Ideally, 
one needs to know the 4-momentum of the parton when it has been produced in the hard scattering and 
after it has been modified by the medium.
This can be achieved by selecting particular hard processes in which there is a photon in the final state.
Since the photon does not interact with the medium, its 4-momentum is not modified and thus provides a measure of the hard scattered parton emitted back-to-back with the photon.
Measuring the hadrons opposite to the photon  
is thus a promising way to measure the jet fragmentation and misalignment between photon and hadrons to quantify the modifications due to the medium. 

\section{Approach Validation with Monte-Carlo Data}
The experimental technique consists in tagging events with a well
identified high energy direct photon and measuring the distribution
of hadrons emitted oppositely to the photon as a function of the
parameter $x_{E} =-\vec{p}_{T}^{h}\cdot \vec{p}_{T}^{\gamma}/\mid
p_{T}^{\gamma}\mid ^{2}$. Such a measurement requires an excellent
direct photon identification and the measurement of charged and
neutral hadrons with good $p_{T}$ resolution. In ALICE, the
electromagnetic calorimeters, PHOS ($|\Delta\eta|<0.12$ and $\Delta
\phi$ =100$^o$) and EMCal ($|\Delta\eta|<0.7$ and $\Delta \phi$
=100$^o$)~\cite{PHOS, EMCal}, are capable to measure photons with high
efficiency and resolution~\cite{PID}. The
central tracking system (ITS and TPC), covers the pseudorapidity
$-0.9 \leq \eta \leq +0.9$ and the full azimuth, is helpful for direct photon extraction with the isolation technique. 

We have first established the feasibility of $\gamma$--hadrons
correlation measurement with ALICE detectors using Monte-Carlo data.
As a first result~\cite{corr} of this study,  PYTHIA~\cite{pythia} generator is used to simulate $pp$ collisions at $\sqrt{s}~=~14~$TeV containing a  2$\rightarrow$2 process with a direct photon inside PHOS  acceptance. 
we have demonstrated that this
measurement allows us to determine, both the jet fragmentation
distribution and the back-to-back azimuthal alignment of the direct
photon and the jet. However because of the limited acceptance
covered by the calorimeters, the measurement is restricted by
statistics to photon with energies below 50 GeV.
This kinematic region is particularly interesting because
jets of such low energy loose a large fraction of their energy while
traversing the medium, rendering the medium modification most
visible. In addition, because jets with energy below 50 GeV can
hardly be reconstructed in the heavy-ion environment, the photon
tagging technique provides a sensitive measurement of
jets in this kinematic range. 
Systematic errors due to the improper identification of direct
photons remain, within this kinematic range, lower than statistical
errors from our study~\cite{corr}.

To quantify the medium modification, the photon--hadrons correlation distribution has been studied with events generated in $pp$ collisions at $\sqrt{s}~=~5.5~$TeV containing a  2$\rightarrow$2 process with a direct photon inside EMCal acceptance. They have been generated 
by PYTHIA~\cite{pythia} generator and
qPYTHIA,  which includes a parton energy loss model~\cite{qpythia} with the medium transport parameter $\hat{q}$~=50~GeV$^2$/fm for photon energies between 5 and 200 GeV. 
At this stage of the study,  the heavy-ion collision background has not yet been taken into account.   Direct photons are identified with the isolation technique requiring no hadronic activity around the direct photon candidate inside a given cone size~\cite{Isolation}. Hadrons detected in the azimuthal range $\pi/2 < \Delta \phi < 3\pi/2$ relative to the photon were used to construct the correlation function. The contribution of hadrons from the underlying event was calculated from the hadrons emitted in the same azimuthal hemisphere as the photon.

\begin{figure}[h]
\hspace {4pc}
\begin{minipage}{14pc}
\includegraphics[width=14pc]{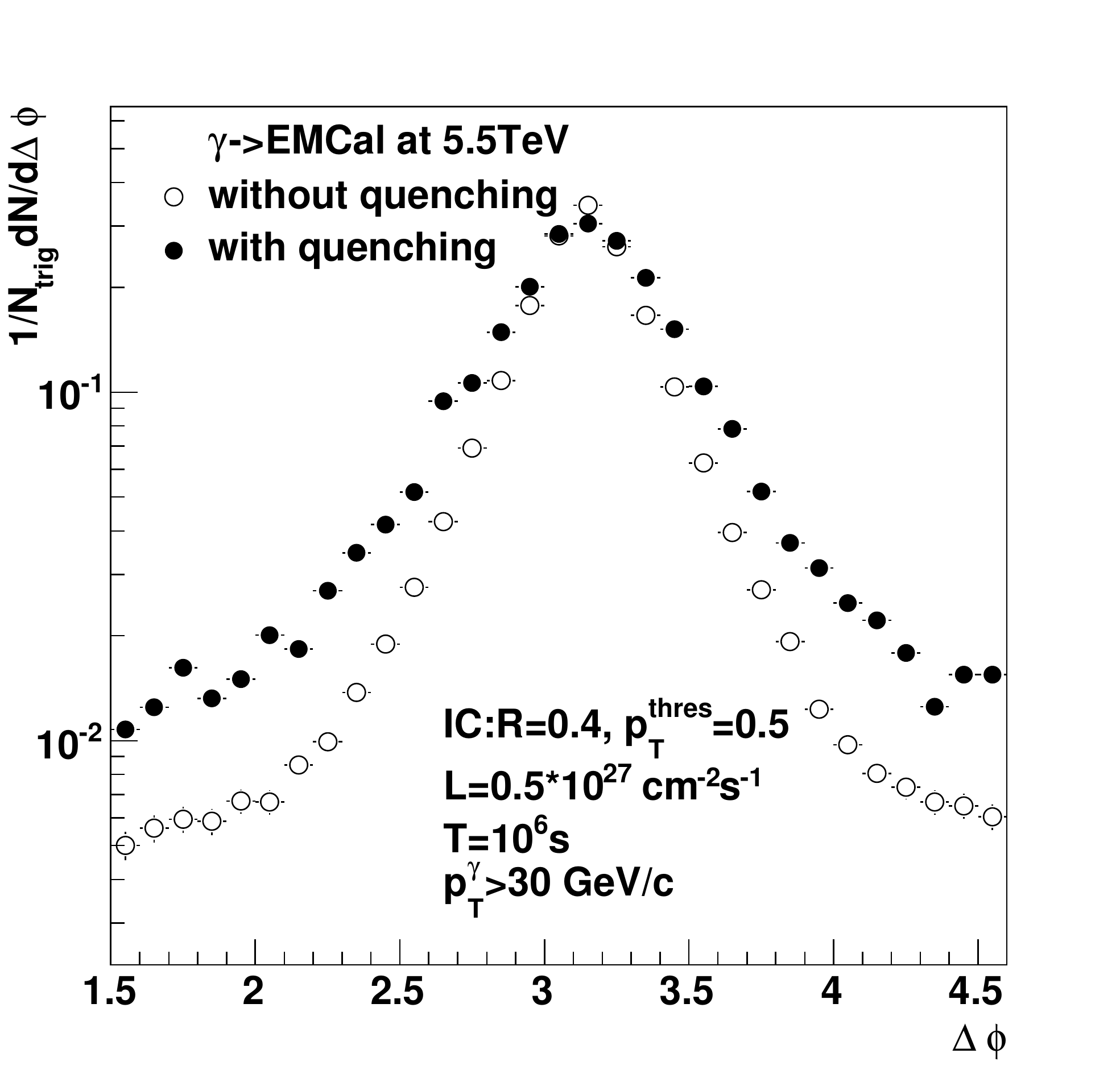}
\caption{\label{fig:DeltaPhi}Relative azimuthal angle distribution $\Delta \phi = \phi_{\gamma}-\phi_{hadron}$ for $\gamma$-jet events in pp collisions at $\sqrt{s}$~=~5.5~TeV. }
\end{minipage}\hspace{2pc}%
\begin{minipage}{14pc}
\includegraphics[width=14pc]{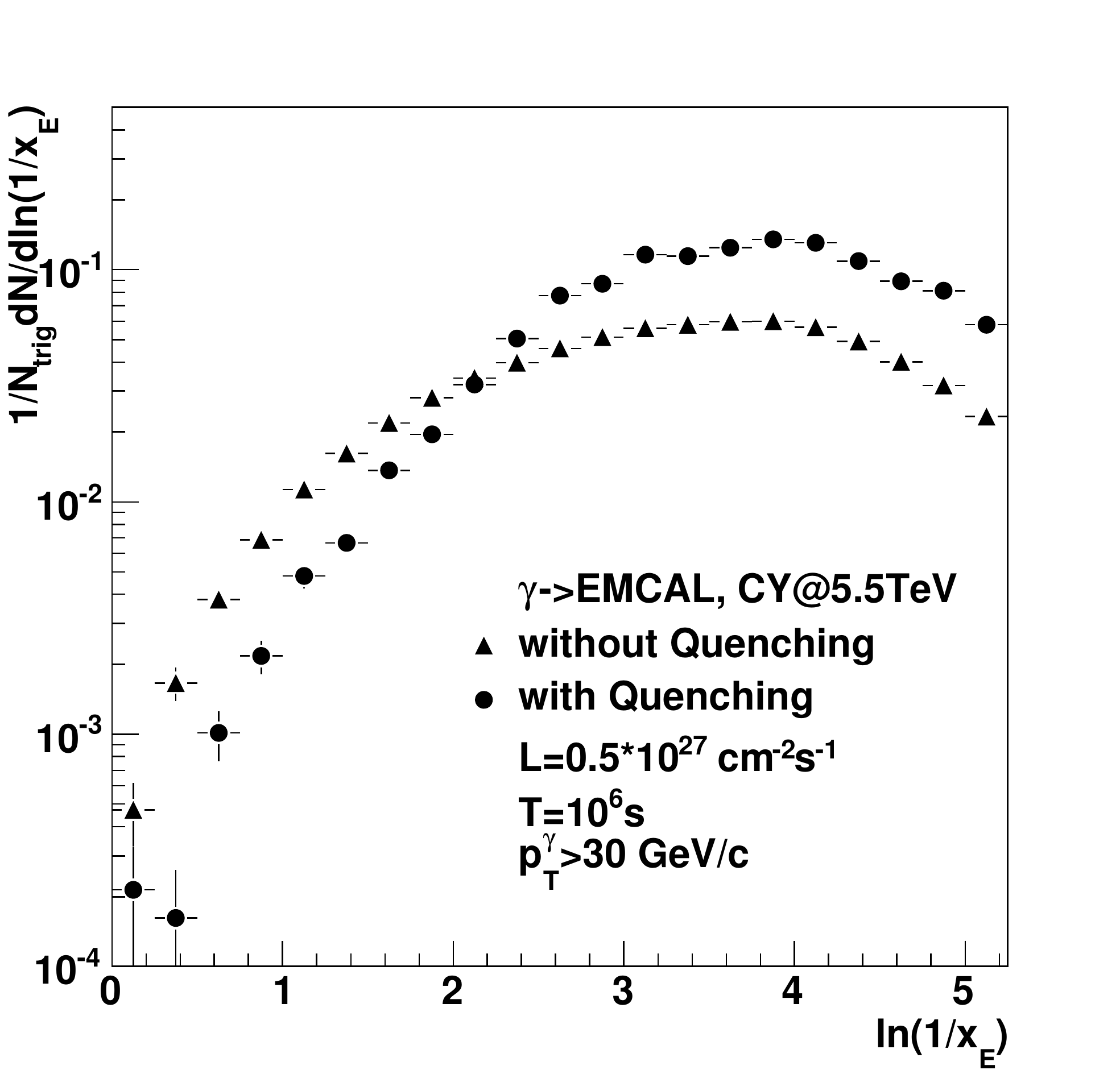}
\caption{\label{fig:Quench}$\gamma$-hadron correlation distributions in
quenched and unquenched PYTHIA events as a function of $ln(1/x_{E})$. }
\end{minipage}
\end{figure}

The relative azimuthal angle, $\Delta \phi=\phi_{\gamma} -\phi_{h} $, between the direct photon and charged hadrons is strongly peaked at $\pi$ as expected for the 2$\rightarrow$2 process (Fig.~\ref{fig:DeltaPhi}).
When medium effects are simulated (qPYTHIA), the $\Delta \phi$ distribution becomes broader. The broadening can be related to the medium transport parameter $\hat{q}$.
However, the effect is quite small which will make the measurement in the heavy-ion environment quite challenging.
A stronger signal is expected to be observed in  the photon--hadrons distribution from heavy-ion collisions when compared to the distribution from pp collisions.
The resulting photon-triggered hadrons distributions, after subtraction of underlying events, are shown in Fig.~\ref{fig:Quench}, normalized to the number of trigger particles found in corresponding generation.
The statistical errors are estimated from the annual yield of photon events with $p_{T}$ larger than 30 GeV we anticipate to collect during one PbPb run at nominal luminosity~\cite{ALICE}.
The distribution exhibits the expected suppression at high $x_{E}$, due to the enegy loss of the hard scatered parton and the enhancement at low $x_{E}$ due to the fragmentation of soft gluons radiated in the medium.

\section{Two Particle Correlations in pp@7TeV}
Minimum bias data have been collected in pp collisions at center of mass $\sqrt{s}~=7~$TeV. We have analyzed about 35 million events in a first attempt to measure $\gamma$-hadrons correlations. The trigger particle is selected as the one with the highest transverse momentum measured either in the central tracking system or in the electromagnetic calorimeters. In the calorimeters, electromagnetic particles are detected as clusters of hit calorimeter cells. Roughly we have identified $\pi^{0}$ candidate as a pair of clusters which invariant mass matches the $\pi^{0}$ mass range, $135\pm 15$~MeV, and single clusters (which do no pair with another cluster) as direct photon candidates. No particle identification has been applied yet so that the single cluster sample contains a sizable fraction of charged particles which develop a shower in the calorimeters.

\begin{figure}[h]
\hspace {4pc}
\begin{minipage}{14pc}
\includegraphics[width=15pc]{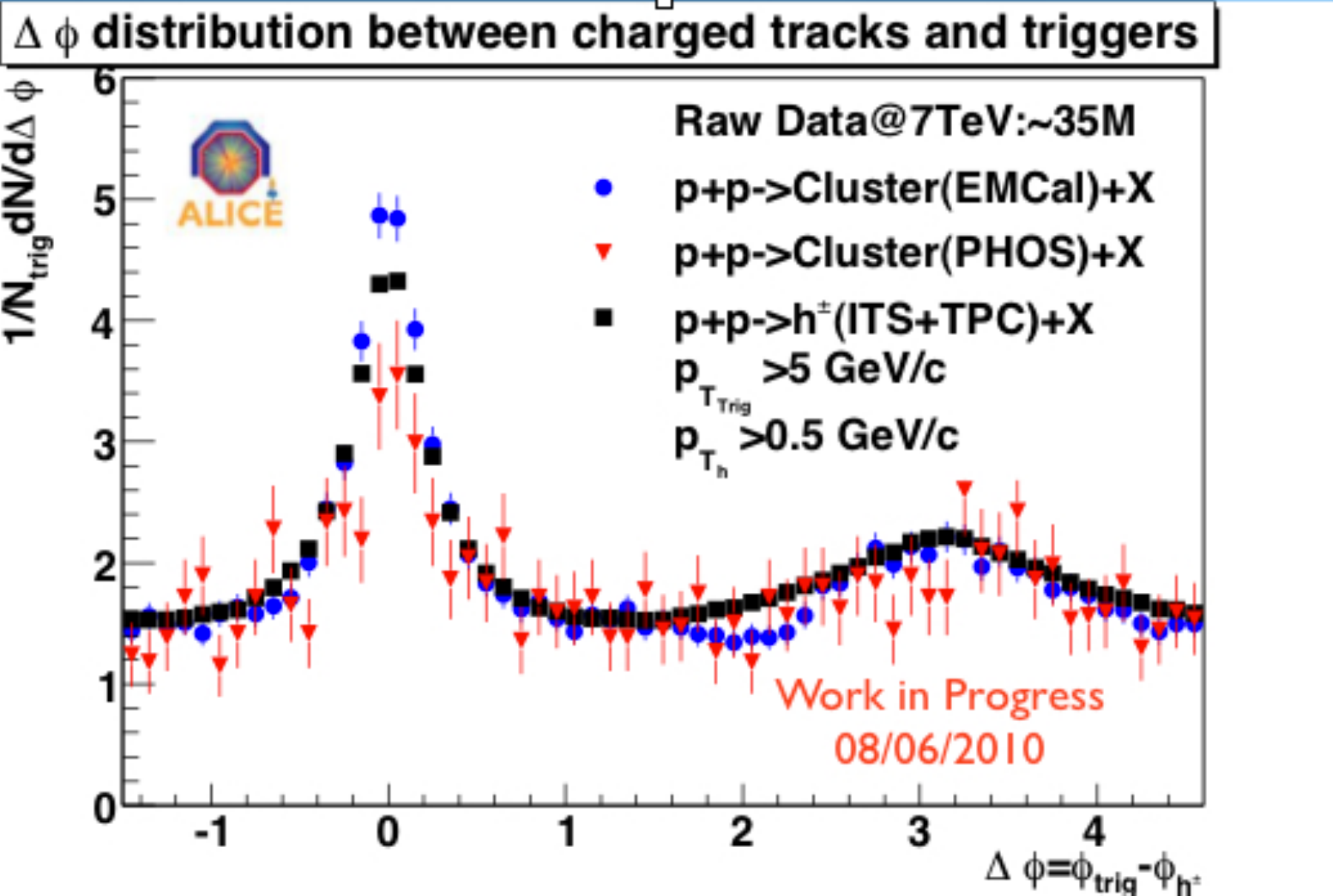}
\caption{\label{fig:pp7Dphi}Relative azimuthal angle distribution $\Delta \phi = \phi_{trigger}-\phi_{hadron}$ in pp collisions at $\sqrt{s}$~=~7~TeV. }
\end{minipage}\hspace{2pc}%
\begin{minipage}{14pc}
\includegraphics[width=15pc]{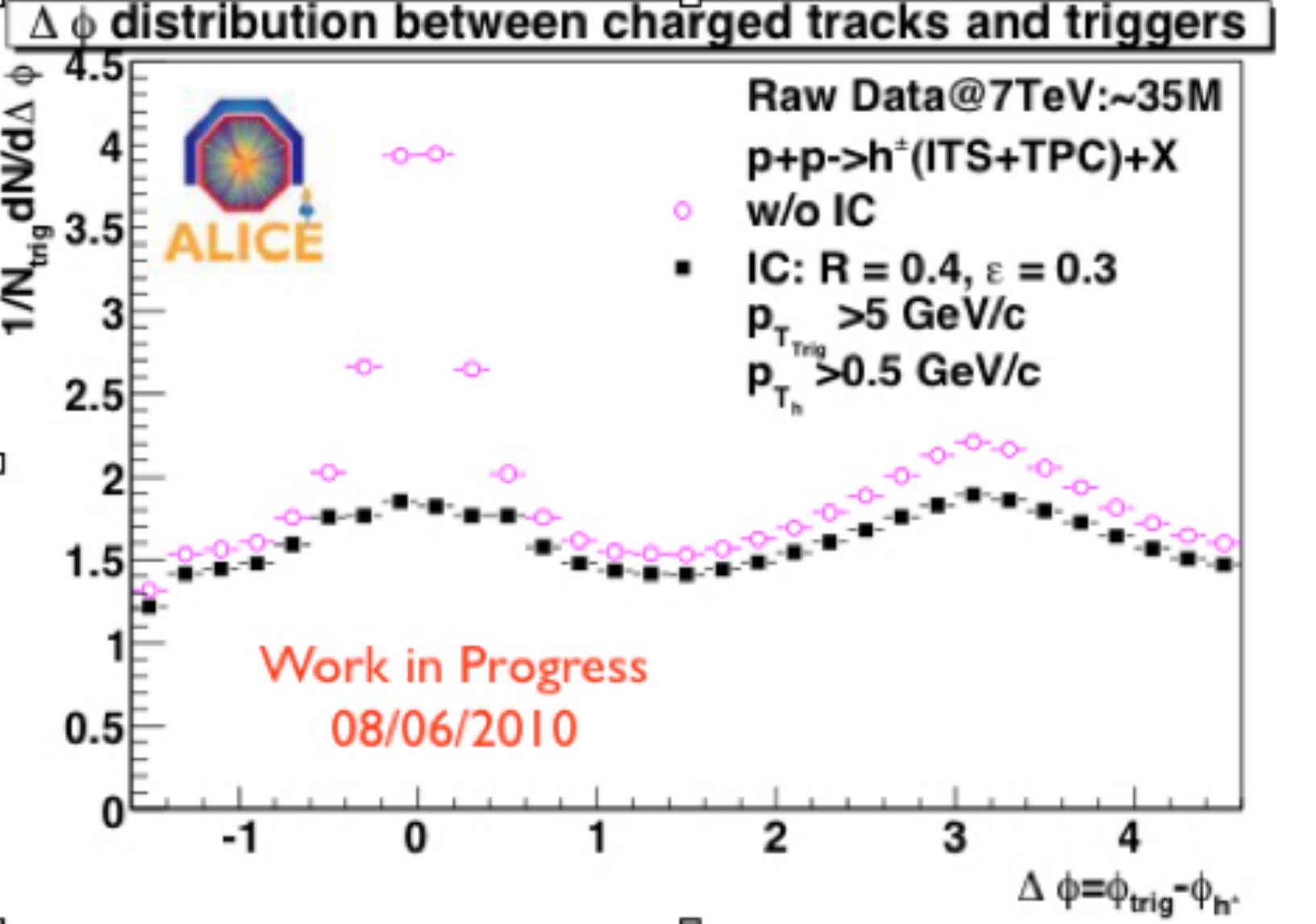}
\caption{\label{fig:IsoTrigDphi}Azimuthal correlation distributions before and after isolation cut (IC) on the trigger particles with $p_{T}  > 5GeV/c$}
\end{minipage}
\end{figure}

The azimuthal correlation between the trigger particle (charged particle, $\pi^0$ candidate, single cluster) and the charged hadrons are shown in Fig.~\ref{fig:pp7Dphi}. The near side ($\Delta\phi=0$) and away side ($\Delta \phi=\pi$) peak are clearly observed. Note, however, that these distributions have not been corrected for efficiency. It is interesting to remark that at this very preliminary stage of the analysis we find that  the underlying event background level, outside the peaks region, is independent of the type of trigger particles, giving some confidence in the measurement. By applying an isolation selection on the trigger candidate, where hadron activity carries less than 30~\% transverse momentum of the trigger candidate inside a cone with size $R = 0.4$ required,
the probability of direct photon or single particle jets in the sample enhances. Comparing the azimuthal correlation with and without isolation selection,
obviously, a suppression of the near side peak is observed, but the away side peak is almost unaffected, as expected (Fig.~\ref{fig:IsoTrigDphi}).
However this preliminary analysis does not allow to 
draw any conclusion other that these results indicate the expected behaviour.
The isolation parameters are not well adjusted and especially in our case only charged tracks are considered in our isolation cone due to the limited calorimeter acceptance (40~\% EMCAL and 60~\% PHOS have been installed so far).

\section{Summary and outlook}
The feasibility to measure $\gamma$--hadrons correlation in pp collisions and
medium modification effect in PbPb collisions with ALICE has been evaluated. 
Such a measurement  provides an exclusive observable sensitive to the properties of the medium formed in heavy-ion collisions. 
So far only a preliminary analysis has been performed
on a small fraction of the data collected by ALICE. Exciting physics
will certainly come with the final analysis of large statistics with
well calibrated detectors.
\section*{Acknowledgement}
The work is partially supported by the NSFC (10875051, 10635020 and
10975061), the Key Project of Chinese Ministry of Education
(306022), the Program of Introducing Talents of Discipline to
Universities of China (QLPL200909, B08033 and CCNU09C01002).

\section{References}

\end{document}